# Extra phase noise from thermal fluctuations in nonlinear optical crystals

J. E. S. César[1], A. S. Coelho[1], K. N. Cassemiro[2], A. S. Villar[3], M. Lassen[3], P. Nussenzveig[1], and M. Martinelli[1]

[1] *Instituto de Física, Universidade de São Paulo,*
*Caixa Postal 66318, 05315-970 São Paulo, SP, Brazil*
[2] *Max Planck Junior Research Group Günther Scharowskystr. 1 / Bau 24 91058 Erlangen, Germany.*
[3] *Max Planck Institute for the Science of Light,*
*University of Erlangen-Nuremberg, Staudtstr. 7/B2, 91058 Erlangen, Germany*
(Dated: June 22, 2009)

We show theoretically and experimentally that scattered light by thermal phonons inside a second-order nonlinear crystal is the source of additional phase noise observed in Optical Parametric Oscillators. This additional phase noise reduces the quantum correlations and has hitherto hindered the direct production of multipartite entanglement in a single nonlinear optical system. We cooled the nonlinear crystal and observed a reduction of the extra noise. Our treatment of this noise can be successfully applied to different systems in the literature.

In quantum information, processes with continuous variables have attracted growing attention, involving the manipulation of states of light and atoms. Examples are the unconditional quantum teleportation of states of the electromagnetic field [1], and teleportation and storage of quantum information between light and atoms [2]. Many of these experiments involved nonclassical states of the electromagnetic field, either squeezed states or N-partite entangled states, generally obtained from the combination of squeezed states in a set of beam-splitters [3]. Quadrature squeezed states of the electromagnetic field attracted renewed interest, after the initial promises of ultra-sensitive spectroscopy and noiseless communication channels, predicted in the late 80's.

A source of nonclassical states are nonlinear crystals inside optical cavities. This was the case for the first generation of squeezing in parametric down conversion [4], soon followed by the observation of quantum correlations between separate fields [5] and production of entangled states [6] in an Optical Parametric Oscillator (OPO), and squeezing in Second Harmonic Generation (SHG) [7]. At the heart of this process is the parametric coupling of three fields, used for twin photon generation, but with the enhancement and mode selection given by a resonant cavity. Early studies [8] theoretically demonstrated that a nonlinear crystal inside a cavity would produce squeezing in different modes, and entanglement between pairs of modes.

Despite this long list of successful implementations showing excellent agreement with the theory, the measurement of other nonclassical fundamental characteristic features remained elusive. Phase quadrature squeezing of the pump field in the above-threshold OPO [9] presented an unexplained excess noise, that could not be accounted for by any known source in the system. Furthermore, the first measurements of noise correlations between the phases of signal and idler beams in OPO's above threshold resulted in great excess noise where squeezing was expected [10]. The first measurement of this quantum correlation, demonstrating entanglement of signal and idler beams [11], was limited to operation very close to threshold. An increase in the pump power rapidly degraded the quantum phase correlations. This was one of the main reasons why signal and idler entanglement remained unobserved for almost 20 years after its prediction [8]. Measurements performed later by other groups [12, 13] confirmed this excess noise, and many attempts where made to explain its origin [14–16]. The same kind of problems were observed in the coupling of infrared and green fields in a degenerate OPO/SHG [17]. Moreover, the direct production of three mode entanglement was also predicted [18], but the first experimental implementations could not overcome this unknown source of extra phase noise [19].

Here, we make a detailed treatment of this extra noise in the OPO and experimentally verify its predictions. The model is described in section I, and the expected behavior for an OPO is then presented. A full characterization of this noise in the three modes of an optical cavity encompassing a nonlinear crystal is given in section II. This complete description of the noise presents good agreement with the experimental observations of the covariance matrix, measured for the OPO described in ref. [19] (section III). We show that this additional noise is consistent with a model of incoherent (thermal) phonon noise. Random density fluctuations are induced and, via the change of the refractive index, entail phase fluctuations as described in section IV. We also apply our model to other results reported in the literature, with good agreement. Finaly, we conclude by a discussion on proposals for reducing or eliminating this noise in experiments for the production of squeezed or entangled states involving phase quadratures.

## I. MODEL FOR THE INTRODUCED PHASE NOISE

In the present study, we will consider that the crystal, although homogeneous, can have small fluctuations in its permittivity. The refractive index fluctuates owing to local density fluctuations associated with acoustic phonons inside the crystal. This density change results in Stokes and Brillouin light scattering [20] with frequency shifts in the scattered light. In the present case, we will be interested in the fraction of the scattering that is coupled into the cavity modes, with small shifts in the frequency (within the cavity bandwidth).

If we consider that the constitutive relation between the displacement vector $\boldsymbol{D}$ and the electric field $\boldsymbol{E}$ includes a random local fluctuation in permittivity $\delta\boldsymbol{\varepsilon}$

$$\vec{D} = \varepsilon\vec{E} + \delta\varepsilon(z,t)\vec{E}, \quad (1)$$

we have, from the Maxwell equations

$$\nabla \times \vec{E} = -\frac{\partial}{\partial t}\vec{B}, \text{ and } \nabla \times \vec{H} = \frac{\partial}{\partial t}\vec{D}, \quad (2)$$

the resulting wave equation coupling this fluctuation to the propagating field is thus given by

$$\nabla^2 \vec{E} = \mu_0 \varepsilon \frac{\partial^2}{\partial t^2}\vec{E} + \mu_0 \frac{\partial^2}{\partial t^2}(\delta\varepsilon \vec{E}). \quad (3)$$

In the present case, we will be interested in a single frequency, of a given polarization, referring to a mode $j$ of the optical cavity. We can thus separate a slowly varying amplitude of the field $A_j$ from the fast oscillating part

$$E_j(\vec{r},t) = Re[A_j(\vec{r},t)\,exp(in_jk_jz - i\omega_jt)], \quad (4)$$

with $k_j = \omega_j/c$, where $c$ is the speed of light in vacuum and $n_j = \sqrt{\varepsilon_j/\varepsilon_0}$ is the refractive index. The amplitude includes both the mean value $\bar{A}_j$ and the fluctuations induced by the permittivity:

$$A_j(\vec{r},t) = \bar{A}_j(\vec{r}) + \delta A_j(\vec{r},t). \quad (5)$$

Since we are dealing with the resonant modes in a cavity, the solutions of the wave equation can be limited to the paraxial approximation. The amplitude of the field will satisfy the paraxial wave equation [21], so $\bar{A}_j(\vec{r}) = \alpha_j u_j^{(h)}(\vec{r})$, where the transverse mode $(h)$ for the longitudinal mode $j$ can be described in the Hermite or Laguerre-Gauss basis by the function $u_j^{(h)}(\vec{r})$. The mode amplitude $\alpha_j$ is treated as a constant, an approximation which is valid in the case that field depletion is small.

The wave equation for the fluctuating part can then be solved, in the paraxial approximation, considering a slowly varying envelope for the field fluctuation. Examining the contribution of scattering at the sideband of the carrier we have, in the frequency domain,

$$\delta A_j(\vec{r},\Omega) = i\frac{n_jk_j}{2\varepsilon_j}\alpha_j u_j^{(h)}(\vec{r})\delta\varepsilon_j(\vec{r},\Omega)\delta z, \quad (6)$$

for a given position of the wavefront. Note that this contribution occurs in quadrature with the local field mean amplitude $\alpha_j u_j^{(h)}(\vec{r})$.

The total contribution of the fluctuation for a given field mode, $\delta\alpha_j(z,\Omega)$, can be evaluated using the orthogonality of the basis states ($\int u_j^{(h)}(\vec{r})u_j^{(g)*}(\vec{r})\,dx\,dy = \delta_{hg}$). Performing the transverse integration of the fluctuating part of eq. 5, with the weight factor $u_j^{(h)*}(\vec{r})$ accounting for the coupling of the scattered field into the cavity mode, we have

$$\delta\alpha_j(z,\Omega) = \int \delta A_j(\vec{r},\Omega)u_j^{(h)*}(\vec{r})\,dx\,dy. \quad (7)$$

Finally, the total contribution of the fluctuation will include the integration over the crystal length. The resulting added noise will be a fluctuation term, named $\delta Q_j$, to be added in the stochastic treatment of the quantum fluctuations inside the cavity. This term is in quadrature with the amplitude $\alpha_j$ of the field. Thus, it can be described as an additional phase noise for an intense mean field.

$$\begin{aligned}\delta Q_j(\Omega) &= -i\int \delta\alpha_j(z,\Omega)dz \quad (8)\\ &= \frac{n_jk_j}{2\varepsilon_j}\alpha_j\int |u_j^{(h)}(\vec{r})|^2\delta\varepsilon_j(\vec{r},\Omega)\,dx\,dy\,dz.\end{aligned}$$

As we can see, the random fluctuation in the crystal permittivity will not produce any significant contribution to the amplitude quadrature. The resulting change in the refractive index will only produce phase fluctuations that couple to the resonant cavity mode, within the cavity bandwidth, resulting in an additional source of noise. This noise will be particularly important for experiments in quantum optics, as we observe in the next section, when we add this term to the treatment of quantum fluctuations in the OPO.

### A. Excess Noise in the OPO

Most of the systems involving second order nonlinearities inside a cavity can be described by the evolution of the three coupled fields. Let us define our amplitude and phase quadrature operators from the creation operator of the field $\hat{a} = e^{i\theta}(\hat{p} + i\hat{q})$, where the arbitrary phase $\theta$ is chosen such that $\langle\hat{q}\rangle = 0$. With this choice of $\theta$, quadrature $\hat{p}$ is associated with the amplitude, and $\hat{q}$ is associated with the phase in the case of an intense field. In a linearized description of the fields, with $\delta\hat{a} = \hat{a} - \langle\hat{a}\rangle$, field fluctuations are given by the vector

$$\vec{X} = [\delta\hat{p}_0, \delta\hat{q}_0, \delta\hat{p}_1, \delta\hat{q}_1, \delta\hat{p}_2, \delta\hat{q}_2]^T. \quad (9)$$

The operators can be replaced by c-numbers using the master equation for the density operator and a quasi-probability representation of the field, such as the Wigner distribution, to obtain a Fokker-Planck equation [22] and its equivalent description by Langevin equations. In this linearized approach, equations for the evolution of field fluctuations inside a cavity are described by the following Langevin equation

$$\tau \frac{\partial}{\partial t}\vec{X} = \boldsymbol{M_A}\vec{X} + \boldsymbol{M_\gamma}\vec{X}_1^{in} + \boldsymbol{M_\mu}\vec{X}_2^{in} + \vec{Q}, \quad (10)$$

where $\tau$ is the round trip time of the wave inside this cavity. The drift matrix $\boldsymbol{M_A}$ describes the evolution of the field fluctuations in a round trip, including attenuation, parametric amplification and phase shifts. The next two terms couple the input field fluctuations to the cavity through the input coupler ($\boldsymbol{M_\gamma}$) and the spurious losses ($\boldsymbol{M_\mu}$), and are associated to the diffusion in a Langevin process. This coupling is described by the diagonal matrices

$$\boldsymbol{M_\gamma} = diag\left[\sqrt{2\gamma_0}, \sqrt{2\gamma_0}, \sqrt{2\gamma_1}, \sqrt{2\gamma_1}, \sqrt{2\gamma_2}, \sqrt{2\gamma_2}\right],$$
and
$$\boldsymbol{M_\mu} = diag\left[\sqrt{2\mu_0}, \sqrt{2\mu_0}, \sqrt{2\mu_1}, \sqrt{2\mu_1}, \sqrt{2\mu_2}, \sqrt{2\mu_2}\right], \quad (11)$$

connecting the input fields $\vec{X}_1^{in}$ and vacuum fluctuations $\vec{X}_2^{in}$ through the input ports. In this description, spurious losses in each mode $j$ are given by $2\mu_j$, and the mirror transmission is given by $T_j = 2\gamma_j$.

The excess phase noise coming from the phonons in the crystal is given by the additional contribution to the phase fluctuations (eq. 8). This stochastic fluctuation will be present in the three relevant cavity modes, described by the vector

$$\vec{Q} = [0, \delta Q_0, 0, \delta Q_1, 0, \delta Q_2]^T. \quad (12)$$

In the frequency domain, eq. 10 can be described by $i\Omega\vec{X} = \boldsymbol{M_A}\vec{X} + \boldsymbol{M_\gamma}\vec{X}_1^{in} + \boldsymbol{M_\mu}\vec{X}_2^{in} + \vec{Q}$, and the intracavity fluctuations will then be given by

$$\vec{X}(\Omega) = [i\Omega\boldsymbol{I} - \boldsymbol{M_A}]^{-1}(\boldsymbol{M_\gamma}\vec{X}_1^{in} + \boldsymbol{M_\mu}\vec{X}_2^{in} + \vec{Q}). \quad (13)$$

Using the input-output formalism [23] we have the fluctuations of the output port of the cavity given by the reflection of the input, added to the transmitted internal fluctuation: $\vec{X}^{out}(\Omega) = \boldsymbol{M_\gamma}\vec{X}(\Omega) - \vec{X}_1^{in}$. In this case we can easily calculate the covariance matrix of the output field $\boldsymbol{V} = \vec{X}^{out}(\Omega)[\vec{X}^{out}(-\Omega)]^T$, resulting in

$$\boldsymbol{V} = \boldsymbol{I} + \boldsymbol{V}_{pure} + \boldsymbol{V}_{loss} + \boldsymbol{V}_{phase}. \quad (14)$$

The identity matrix $\boldsymbol{I}$ is associated with the Standard Quantum Level (SQL) of noise, characteristic of coherent states (including vacuum). The next term

$$\boldsymbol{V}_{pure} = \quad (15)$$
$$\boldsymbol{M_\gamma}[i\Omega\boldsymbol{I} - \boldsymbol{M_A}]^{-1}\boldsymbol{M_\gamma}\boldsymbol{M_\gamma}\{[-i\Omega\boldsymbol{I} - \boldsymbol{M_A}]^{-1}\}^T\boldsymbol{M_\gamma}$$
$$-\boldsymbol{M_\gamma}\left([i\Omega\boldsymbol{I} - \boldsymbol{M_A}]^{-1} + \{[-i\Omega\boldsymbol{I} - \boldsymbol{M_A}]^{-1}\}^T\right)\boldsymbol{M_\gamma}$$

stands for the squeezing and the quantum correlation leading to entanglement in this system. For unitary intracavity operations, the resulting state will be pure and $Det[\boldsymbol{I} + \boldsymbol{V}_{pure}] = 1$ [24]. There is also a noise term owing to the input fluctuations through the spurious losses, uncorrelated with the reflected fluctuations through the input port, degrading the level of squeezing and/or entanglement and state purity as well:

$$\boldsymbol{V}_{loss} = \quad (16)$$
$$\boldsymbol{M_\gamma}[i\Omega\boldsymbol{I} - \boldsymbol{M_A}]^{-1}\boldsymbol{M_\mu}\boldsymbol{M_\mu}\{[-i\Omega\boldsymbol{I} - \boldsymbol{M_A}]^{-1}\}^T\boldsymbol{M_\gamma}.$$

Finally, the internal added noise will depend on the covariance matrix for the additional phase fluctuations $\boldsymbol{V_Q}(\Omega) = \vec{Q}(\Omega)\vec{Q}^T(-\Omega)$, affected by the cavity interaction and the nonlinear coupling given by the drift matrix $\boldsymbol{M_A}$:

$$\boldsymbol{V}_{phase} = \quad (17)$$
$$\boldsymbol{M_\gamma}[i\Omega\boldsymbol{I} - \boldsymbol{M_A}]^{-1}\boldsymbol{V}_Q(\Omega)\{[-i\Omega\boldsymbol{I} - \boldsymbol{M_A}]^{-1}\}^T\boldsymbol{M_\gamma}.$$

A detailed evaluation of the terms in the covariance matrix $\boldsymbol{V_Q}$ will be given in the section IV. For the moment, considering the fluctuation term presented in eq. 8, we expect that the covariance terms of this extra noise are given by:

$$\langle\delta Q_j(\Omega)\delta Q_k(-\Omega)\rangle = \frac{n_j k_j n_k k_k}{4\varepsilon_j\varepsilon_k}\sigma_{jk}\alpha_j\alpha_k$$
$$= \eta_{jk}\sqrt{P_j P_k}. \quad (18)$$

The noise coupling term $\eta_{jk}$ depends on the wavelength through $k_j$ and $k_k$ and on the evaluation of the double integration coming from the product of eq. 8 for two different modes, resulting in $\sigma_{jk}$. Here we have used the fact that the intracavity amplitude of the field $\alpha_j$ is proportional to the square root of the intracavity power $P_j$

From this simple model, a wide variety of experiments involving two or three fields coupled by nonlinear effects can be studied, implying in different forms of matrix $\boldsymbol{M_A}$ in eq. 10. In the case that signal and idler modes are degenerate, as in a degenerate OPO, seeded OPA, subthreshold OPO and SHG, the problem can be reduced to the evaluation of a matrix of size $4 \times 4$. In the case of a cavity without nonlinear interaction, the drift matrix is reduced to $\boldsymbol{M_A} = -(\boldsymbol{M_\gamma}^2 + \boldsymbol{M_\mu}^2)/2$. For a non-degenerate, unseeded OPO above threshold, with balanced losses for the signal and idler modes ($\gamma_1 = \gamma_2 = \gamma$ and $\mu_1 = \mu_2 = \mu$), the drift matrix has a simple form in the case of exact resonance (zero detuning) [25]:

$$\boldsymbol{M_A} = \begin{bmatrix} -\gamma_0' & 0 & -\gamma'\beta & 0 & -\gamma'\beta & 0 \\ 0 & -\gamma_0' & 0 & -\gamma'\beta & 0 & -\gamma'\beta \\ \gamma'\beta & 0 & -\gamma' & 0 & \gamma' & 0 \\ 0 & \gamma'\beta & 0 & -\gamma' & 0 & -\gamma' \\ \gamma'\beta & 0 & \gamma' & 0 & -\gamma' & 0 \\ 0 & \gamma'\beta & 0 & -\gamma' & 0 & -\gamma' \end{bmatrix}, \quad (19)$$

with the coeficient $\beta$ being given by the ratio of the amplitude for signal and idler fields $\alpha$ by the pump field

amplitude $\alpha_0$:

$$\beta = \frac{\alpha}{\alpha_0} = \sqrt{\frac{\gamma_0}{\gamma}}\sqrt{\sqrt{\frac{P_{in}}{P_{th}}} - 1}, \quad (20)$$

where $P_{in}$ is the input power of the pump beam, and $P_{th}$ is the oscillation threshold. $2\gamma'_0 = 2(\gamma_0 + \mu_0)$ and $2\gamma' = 2(\gamma + \mu)$ are the sum of round trip losses.

Before we apply this model to the OPO described in ref. [19], we will characterize the coupling terms $\eta_{jk}$ of eq. 18, describing the dependence of the noise matrix $\mathbf{V_Q}$ with the intracavity power for each mode. This can be done by injection of a classical, coherent field in each mode of the cavity involved in the parametric oscillation, and studying the dependence of the phase noise of the output field with the intracavity power (proportional to the incident power).

## II. EXPERIMENTAL CHARACTERIZATION OF THE EXCESS NOISE

We investigated the spurious excess noise using the experimental system as described in [19], reproduced here in fig. 1. The cavity is pumped by the second harmonic of a Nd:YAG laser (Innolight Diabolo), filtered with a mode cleaning cavity to ensure that pump fluctuations are reduced to the SQL for frequencies above 20 MHz. The filter cavity bandwidth is 2.3 MHz, and its transmission is 55% for the carrier frequency. The filtered pump beam is then injected in the OPO, with adjustable power, through the input coupler (IC) with reflectivity of 70% for the pump field (532 nm). The reflected pumped field is recovered from the Faraday rotator (FR). The infrared output coupler (OC) has a reflectivity of 96% at $\approx 1064$ nm. Both mirrors are deposited on concave substrates with curvature radius of 50 mm. The crystal is a bulk, High Gray-Tracking Resistant KTP (HGTR-KTP, Potassium Titanyl Phosphate, by Raicol Crystals) cut for type-II phase matching, with length $\ell = 12$ mm, and anti-reflective coating for both wavelengths. The average free spectral range for the three modes is 5.1(1) GHz. The cavity finesses for pump, signal, and idler modes (the latter defined as the mode with the same polarization as the pump) are, respectively, 16, 135, and 115. The overall detection efficencies are 87% for the infrared beams and 65% for the pump, accounting for detector efficiencies and losses in the beam paths. An infrared beam could be injected in the OPO cavity using an harmonic beam splitter (HS). Its polarization could be adjusted with a half-wave plate (not shown), to match each mode of the cavity with the birefringent crystal inserted.

In this setup, the initial study of the excess noise produced inside the cavity was done by the injection of an intense field in each one of the modes involved in the OPO. In all the measurements presented here, care was taken to avoid phase matching for the relevant parametric processes of up and downconversion. The absence of stimulated parametric conversion was assured by keeping the pump beam below the OPO threshold. For the infrared injection, second harmonic generation is avoided by careful temperature tuning of the crystal. All measurements were made keeping the OPO at exact resonance by a simple dither-and-lock technique applied to the cavity lenght using a piezo actuator on the input coupler. Mode 0 (which corresponds to the pump mode in the OPO) is investigated by the injection of the 532 nm beam, polarized in the $xy$ plane (referring to crystallographic axes), through the input coupler for the pump mode. The noise and the total power of the reflected field is then measured. Mode 1 (signal mode) is investigated by injection of the 1064 nm beam through the input coupler, which has a very small transmittance at this wavelength ($< 0.2\%$), with polarization parallel to the $z$ axis of the crystal. Mode 2 (idler mode) is investigated with light at the same frequency, with polarization parallel to the $xy$ plane. In both cases, the noise properties of the transmitted light is analyzed.

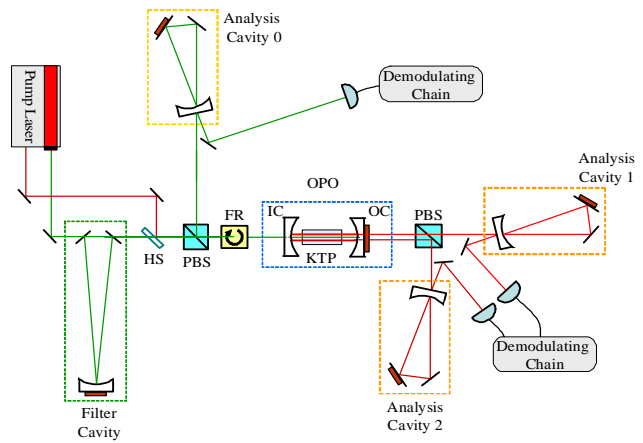

FIG. 1: Setup for the experiment. Pump laser: Innolight Diabolo, HS: harmonic separator/combiner, PBS: polarizing beam splitter. Filtering cavity, analysis cavities and OPO as described in the text.

Phase noise measurements were performed using the ellipse rotation method described in [26, 27], with the help of analysis cavities. Cavities 1 and 2 (for the transmitted infrared beams) have bandwidths of 14(1) MHz, and cavity 0 (for the reflected pump) has a bandwidth of 12(1) MHz. This ensures full rotation of the noise ellipse [25] for the chosen analysis frequency of 21 MHz. Mode matching of the beams to the analysis cavities was better than 95%.

An example of such a measurement is presented in fig. 2, where the intensity noise of the beam reflected by the cavity is fitted according to [25], returning the values of the variance of amplitude and phase quadrature noises. Noise was measured using a demodulating chain and an A/D converter, and each measured correla-tion value was an average of 1000 points, acquired with a bandwidth of 600 kHz. SQL calibration was obtained by the measurement of the variance of a coherent field at different intensities, showing a linear dependence with the beam power. Linearity is verified with great precision, such that the shot noise level is determined within 0.5% [28].

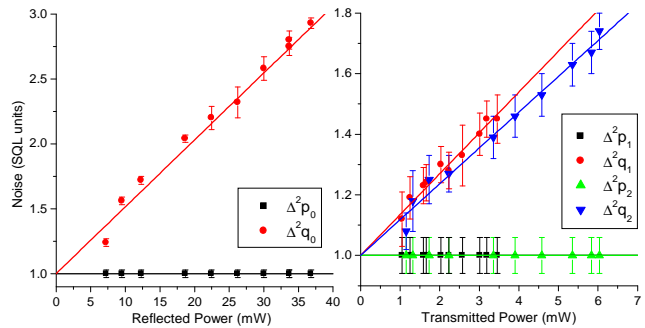

FIG. 3: Measured variances for each mode of the field, as a function of the output power. Left: reflected light from the cavity, for the mode 0 (532 nm). Right: transmitted light for the modes 1 and 2 (1064 nm).

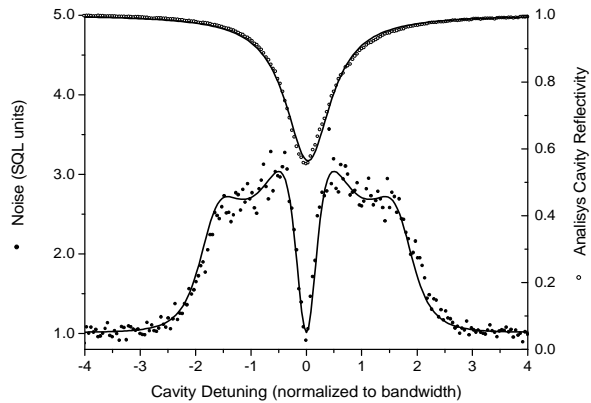

FIG. 2: Measurement of phase noise using the reflection off an empty cavity. The intensity noise of the reflected beam is normalized to the shot noise level, and fitted according to [25], returning the phase and amplitude variances. The reflectivity of the cavity is fitted by a lorenztian

The covariance terms $\Delta^2 Q_k(\Omega)$ from the matrix $V_Q$ and their dependence on the intracavity power are obtained from the excess noise measured from the reflected beam (532 nm) or from the transmitted beam (1064 nm), applying the model of eq. 14 with the drift matrix $M_A$ for uncoupled fields. The measured variance for amplitude and phase quadratures of mode 0 is shown in fig. 3 (left) for different values of the reflected power. Amplitude fluctuations remain at the shot noise level and the phase quadrature presents an excess noise which varies linearly with the power, consistent with the results observed in [14]. Similar results were obtained for modes 1 and 2 (fig. 3 - right). The small transmission of the input coupler provided a strong attenuation of the IR beam, and the measured results for the cavity transmission without the crystal revealed that the noise level of both quadratures for the transmitted beam were reduced to the standard quantum level. Thus, any excess noise at this wavelength is provided by the crystal and occurs only for the phase quadrature, as expected.

The results agree well with the description of a linear dependence of the internal noise with the light power (from eq. 18, $\Delta^2 Q_j = \eta_{jj} P_j$). Using the measured mirror coupling and losses for this cavity, the noise coupling terms $\eta_{jj}$ relating the added noise to internal circulating power are

$$\eta_{00} = 0.53/W, \eta_{11} = 0.15/W,$$
$$\text{and } \eta_{22} = 0.14/W. \quad (21)$$

Uncertainties of these values are mainly due to their local dependence with the point of the crystal where the beam is injected. Lateral displacements of the crystal resulted in changes of the order of 20%. The first point to notice is the ratio of the coupling constant to the wavelength. We have $\eta_{11} \simeq \eta_{22}$, and $\eta_{00}/\eta_{11} = 3.7(5)$, in good agreement with the dependence expected from eq. 18.

After the characterization of the diagonal terms in the covariance $V_Q$, we studied the cross correlation terms using simultaneous injection of two modes. The crystal temperature is carefully tuned in order to obtain cavity double resonance, while keeping the third mode far from resonance, thereby avoiding system oscillation owing to the injection of a seed in the parametric process.

As a result, the curves plotted in fig. 4 show that the correlation depends on the square root of the power of each beam. Similar results were obtained for the noise correlation between modes 0 and 2, and 1 and 2. From the fitting of these curves we can evaluate the following noise coupling terms

$$\eta_{01} = 0.14/W, \eta_{02} = 0.15/W,$$
$$\text{and } \eta_{12} = 0.087/W. \quad (22)$$

Although the uncertainties are somewhat high (about 20%), the confidence band ensures that correlations are not perfect ($\eta_{jk} < \sqrt{\eta_{jj}\eta_{kk}}$), differently from the treatment in ref. [17].

We conclude that the generated phase quadrature noise depends linearly on the intracavity power, as described in ref. [29], but that the noises of different beams are not perfectly correlated. Moreover, amplitude fluctuations remain uncoupled from this additional noise source. After this characterization, we independently measured all the parameters involved in the evaluation of the covariance matrix of the OPO presented in eq. 14. Now we compare the results obtained from the model, after the parameters' evaluation, with the experimental results.

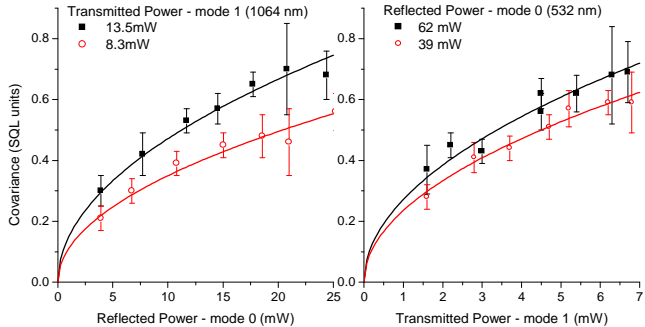
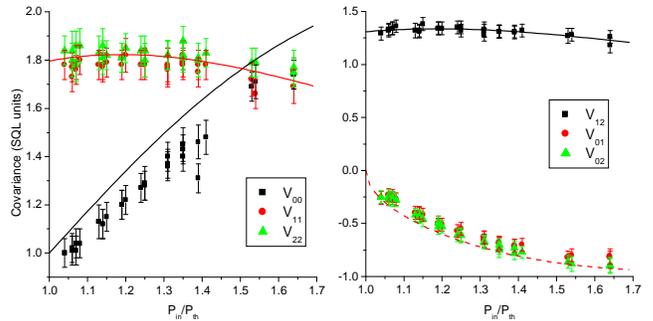

FIG. 4: Left: Covariance of phase quadratures of modes 0 and 1 (orthogonally polarized modes) as a function of the reflected power, for different transmitted power of mode 1. Right: Same, but varying power of mode 1 for two different values of power in mode 0.

FIG. 5: Covariance terms of the amplitude quadratures in the OPO operating above threshold, as a function of the pump power (normalized to threshold). The lines are obtained from the model, without any fitting. Left: Variance for pump (0), signal (1) and idler (2). Right: covariances involving pairs of fields.

## III. NOISE IN THE ABOVE-THRESHOLD OPO

In order to study the noise properties in the OPO, we removed the injected infrared field and increased the power of the green beam, pumping the OPO above threshold. We measured the covariance terms for many different values of input power, ranging from 1 to 1.7 times the threshold power of 70 mW. In the following, we present the experimental data obtained, with its uncertainties, and the expected values from our model. We also include the theoretical predictions in the absence of the additional phonon noise.

According to the model, covariance matrix terms for the amplitude quadratures are unaffected by this additional phase noise, and the results do not change by its inclusion. The results in this case are shown in fig. 5 for the amplitude variances and for the amplitude correlations. The plots are obtained from the model using measured detection efficiency and cavity parameters, without any free parameters. Good agreement is observed, although the amplitude fluctuations of the signal mode are slightly smaller than expected. Noise covariances involving two fields are in very good agreement with theory.

For the phase quadrature, the fluctuations of the refractive index scatter light from the carrier (the mean field amplitude) into the sidebands at 21 MHz. The resulting noise in the output combines the contributions from the parametric process and from this additional noise term. Our results are presented in fig. 6 for the covariances of the phase quadrature fluctuations. The curves are obtained from the values of cavity parameters, and from the measured noise terms given in eqs. 21 and 22 applied to eq. 14. Good agreement is obtained for the variances of signal and idler (fig. 6 - Left). As for the pump, the measured noise level was smaller than expected, but its dependence with the pump power presents good qualitative agreement. The results are, nevertheless, much closer to the the presented model with extra noise than to the OPO model without additional noise (open symbols in the plot), which predicts squeezing for the pump and a nearly constant noise for signal and idler. The agreement for the cross covariance terms is even better (fig. 6 - right), with a very good agreement for the correlations of pump with signal and idler, and a qualitative agreement for signal and idler correlation. The expected values without additional noise fail to describe the observed results. The unbalance of signal and idler correlations with the pump mode can be understood in view of the simplified model for the drift matrix $M_A$, where balanced losses for these modes are used. The good qualitative agreement for these results could be improved by fitting the curves using the noise coupling $\eta_{jk}$ as an adjustable parameter.

Thus, the model allows us to consistently explain the previously unexpected OPO measurements reported in [19]. Correlations among all three fields are now accounted for and we can explain why the predicted pump-signal-idler tripartite entanglement was not observed for these levels of additional noise. Two challenges remain: how can we ascertain that the physical origin of the noise is related to phonons in the crystal and how can we control and reduce it to achieve tripartite entanglement. This is discussed in the following section, where we present a detailed theory for the phonon noise in the crystal and verify experimental signatures of its predictions. Then, we also apply our model to other systems described in the literature, where the same kind of phase quadrature noise behavior was reported.

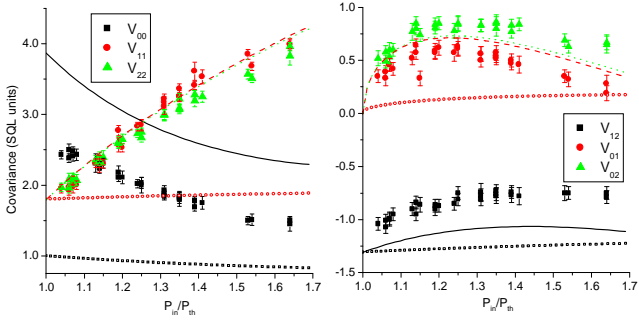

FIG. 6: Covariance of the phase quadratures in the OPO operating above threshold, as a function of the pump power (normalized to threshold). The lines are obtained from the model, without any fitting. Open symbols refer to the model, without additional noise. Left: Variance for pump (0), signal (1) and idler (2). Right: covariances involving pairs of fields.

## IV. PHONON NOISE COUPLING

In order to develop a model that relates the crystal vibration to the phase fluctuations of the fields in their propagation through the crystal, we have to consider the modulation of the permittivity by the local displacement of each part of the crystal. This displacement is represented by a field of local dislocations $\vec{u}(\vec{r})$, where $\vec{r}$ is a position in the crystal. Displacements involving an overall crystal displacement or rotation do not contribute to any change in the crystal energy [30, 31], and therefore cannot affect the dielectric constant. The important quantity, in this case, is the strain

$$S_{lm} = \frac{1}{2}\left[\frac{\partial u_m(\vec{r})}{\partial x_l} + \frac{\partial u_l(\vec{r})}{\partial x_m}\right] \quad (23)$$

that is a six element vector (with $lm = [xx, yy, zz, yz, xz, xy]$).

The strain vector **S** is coupled to the fluctuations of the dielectric tensor $\delta\boldsymbol{\varepsilon}$ by a four rank photoelastic tensor [32]

$$\delta\varepsilon_{id} = -\frac{\varepsilon_i\varepsilon_d}{\varepsilon_0}\sum_{lm} p_{idlm}S_{lm} \ . \quad (24)$$

In the present case, looking for the contribution of the strain to the propagating field, we can restrict this study to the polarization term that is collinear with the electric field, and limit the problem to the case $i = j$. We will therefore keep the short notation $\delta\varepsilon_i = \delta\varepsilon_{ii}$ used in eq. 6.

In order to evaluate the correlation of the phase fluctuations, we can apply eq. 24 to eq. 8, obtaining

$$\langle \delta Q_j(\Omega)\delta Q_k(-\Omega)\rangle = k_j k_k \frac{n_j^3 n_k^3}{4}\alpha_j\alpha_k \times \quad (25)$$

$$\int |u_j(\vec{r})|^2 |u_k(\vec{r'})|^2$$

$$\sum_{(lm),(np)} p_{jjlm}\ p_{kknp}\ \langle S_{lm}(\vec{r},\Omega)S_{np}(\vec{r'},-\Omega)\rangle d\vec{r}d\vec{r'}\ .$$

The integration involves a convolution of the intensity profile of each field by the correlation of the noisy strain vector, implying in an accurate knowledge of this correlation. Nevertheless, further understanding of the scattering process can be obtained if we consider some specific conditions of operation and some assumptions involving this correlation term.

For the case where the beam waist inside the crystal is larger than the coherence length $l_c$ of the acoustic waves of thermal origin, we can approximate the correlation by a delta function

$$\langle S_{lm}(\vec{r},\Omega)S_{np}(\vec{r'},-\Omega)\rangle = S_{lm}^2(\Omega)\delta_{lm,np}\delta(\vec{r}-\vec{r'})l_c^3\ , \quad (26)$$

where $S_{lm}(\Omega)$ is the rms (root mean square) value of the strain fluctuation at a given frequency $\Omega$. In this case, integration of the amplitude envelope of the field in the paraxial approximation is simplified. For fundamental Gaussian modes of waist $w_j(z)$, integration over the crystal volume gives an effective waist for the field $w_{jk}$

$$\int |u_j(\vec{r})|^2|u_k(\vec{r})|^2 d\vec{r} = \int \frac{2}{\pi}\frac{1}{w_j^2(z)+w_k^2(z)}dz = \frac{\ell}{\pi w_{jk}^2}\ . \quad (27)$$

The noise coupling coefficient is inversely proportional to the (effective) area of the beam inside the crystal. Considering that the noise is also proportional to the power, the integration implies in an averaging of the intensity over the crystal volume. The resulting noise is thus proportional to the intensity of the field inside the crystal.

Finally, the product of the photoelectric tensor will result in a coupling term

$$c_{jk}(\Omega) = \sum_{lm} p_{jjlm}p_{kklm}S_{lm}^2(\Omega) \quad (28)$$

obtained by the scalar product of two vectors of six dimensions. If the vectors aren't collinear in this six-dimensional space, we expect to have $c_{jk} < \sqrt{c_{jj}c_{kk}}$. Strain fluctuations will therefore couple differently to each field, resulting in imperfect correlation of the additional phase noise.

Expressing the amplitudes $\alpha_j = \sqrt{P_j/(h\nu_j)}$ in such a way that $|\alpha_j|^2$ is the photon flux per second in the wavefront of the mode, we have

$$\langle \delta Q_j(\Omega)\delta Q_k(-\Omega)\rangle = k_j k_k \frac{n_j^3 n_k^3}{4hc} l_c^3 \; c_{jk}(\Omega) \; \left(\frac{\ell\sqrt{\lambda_j\lambda_k}}{\pi w_{jk}^2}\right) \sqrt{P_j P_k}$$
$$= \eta_{jk}\sqrt{P_j P_k} \,. \quad (29)$$

Some important conclusions can be inferred. Wavelength dependence of the noise coupling $\eta_{jk}$ is restricted to the wavevector modulus $k_j$. The term in parenthesis is equivalent to the ratio of the crystal length to the effective Rayleigh length of the cavity [32]. The cavity geometry determines the beam waist as well, resulting in a noise that is proportional to the intensity. The evaluation of $c_{jk}$ also explains the reason for having $\eta_{jk} < \sqrt{\eta_{jj}\eta_{kk}}$, leading to the expected imperfect noise correlations for the phonon noise contribution to each field: $\langle \delta Q_j(\Omega)\delta Q_k(-\Omega)\rangle < \sqrt{\Delta^2 Q_j(\Omega)\Delta^2 Q_k(\Omega)}$.

As for the coupling $c_{jk}$, its direct evaluation is rather cumbersome, but we can use the fact that the acoustic energy density per unit volume $V$ is related to the strain field as $E/V = \rho v_s^2 S^2/2$ [32], where the crystal density is $\rho$ and $v_s$ is the speed of sound. To first approximation, acoustic phonons can be considered to follow the Bose-Einstein distribution [31] and at temperature T their mean energy is

$$E = \sum_s h\Omega_s \left[\frac{1}{\exp(h\Omega_s/(k_B T))-1} + \frac{1}{2}\right], \quad (30)$$

where different modes $s$ imply in different wavevectors and transverse or longitudinal components. For a given frequency $\Omega \ll k_B T/h$, the dependence of the energy density on the temperature is linear. The coupling term $c_{jk}$ should have a linear behavior with temperature.

Very few parameters involved in the value of $\eta_{jk}$ in eq. 29 can be controlled, in an attempt to reduce the contribution of this noise in quantum optics experiments. Finding crystals with smaller refractive index and photoelastic effects, while maintaining high optical nonlinearities is not a trivial task. On the other hand, cavity parameters (changing the field's spatial profile) and crystal temperature can be changed in order to reduce the extra noise.

### A. Intensity dependence

In order to make the study of the noise dependence on the effective beam waist inside the crystal, we replaced the OPO cavity by a nearly concentric cavity comprising two spherical mirrors of curvature radius $R = 25$ cm, separated by 62.6 mm, resulting in a beam waist of $w_0 = 27.8$ $\mu$m. The resulting Rayleigh length was $z_0' = n\pi w_0^2/\lambda = 8.13(10)$ mm, where $n = 1.788$ is the refractive index for the mode 0 (532 nm, polarized in the $xy$ plane). Input coupler transmittance is $T = 2\gamma = 12.0\%$ and internal losses amount to $2\mu = 3.3\%$. We measured the noise coupling for different crystal positions, obtaining different values for the effective waist $w_{00}$. For different positions $z$ of the crystal relative to the position of the cavity waist, we expect to have, from eq. 27,

$$\frac{\ell\lambda}{\pi w_{00}^2(z)} = n\left\{arctg\left(\frac{2z+\ell}{2z_0'}\right) - arctg\left(\frac{2z-\ell}{2z_0'}\right)\right\}, \quad (31)$$

resulting in the term in parenthesis in eq. 29. As we can see, this term depends only on the cavity geometry (that determines the Rayleigh length), on the crystal length $\ell$, and on the refractive index. The results for the evaluation of the noise coupling at diferent positions inside the cavity are presented in fig. 7. The value of $\eta_{00}$ is fitted to eq. 31 with an overall multiplicative factor as the single adjustable parameter, providing good agreement with the theory.

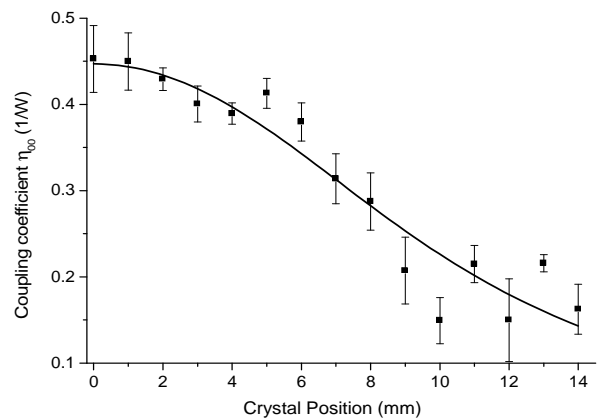

FIG. 7: Noise coupling term $\eta_{00}$ as a function of the position of the crystal inside the cavity. Solid line is a fit to eq. 31.

Therefore, the reduction of the waist inside the crystal increases the noise coupling. On the other hand, it will also reduce the threshold power. Both effects tend to compensate each other. According to eq. 29, the noise level depends on the intensity, as does the parametric amplification [33]. For this reason, changes in the cavity geometry are not expected to significantly affect the contribution of the phonon noise, even though the threshold power is reduced.

### B. Temperature dependence

In order to analyze the temperature dependence of the noise coupling constant $\eta_{00}$, we returned to the original OPO cavity configuration, and controlled the crystal temperature, heating it above room temperature. As discussed, noise power is expected to be proportional to the temperature. Indeed, as can be seen in fig. 8, the results in the range from 257 to 383 K can be adjusted by a linear fit with $\eta_{00}(T) = [5.92(46)10^{-3} \times T/K - 1.38(13)]/W$. This monotonic increase of noise with temperature agrees qualitatively with the naïve model of the phonon noise, although the slope of the curve does not match the parameters in eqs. 29 and 30. Furthermore, it can extrapolated to zero for a finite temperature. This discrepancy can be explained by the fact that the simplified model doesn't account for the temperature dependence of the photoelastic tensor, the crystal density or the speed of sound, and a more complex behavior is expected. Nevertheless, this result shows consistently that cooling the crystal reduces the phonon noise. On the other hand, nonlinear crystals such as $LiNbO_3$, where phase matching is obtained by heating the crystal above $370\,K$, should give an increased contribution from this noise.

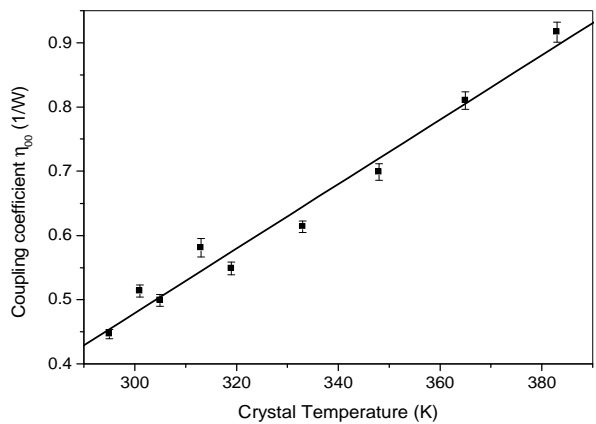

FIG. 8: Temperature dependence of the noise coupling. Data points for temperatures above 300 K were obtained in a single experimental run. For lower temperatures, data points were obtained on different days, leading to the increased dispersion observed. Nevertheless, a clear trend is observed.

## V. APPLICATION TO OTHER EXPERIMENTS

The additional phase noise treated here has been observed in different OPO's by several research groups. In a first attempt to understand its origin [14], we observed excess phase noise in the reflected pump beam, which grew linearly with the incoming pump power below threshold. Since the intracavity pump power above threshold remains constant, this extra noise was initially modeled as a constant noise, similar to what would be expected from a noisy input pump [25].

Peng et al. [15] investigated the effect of pump excess noise in the phase correlation between signal and idler fields as well. Afterwards, they included in their model a gain for the pump phase noise [16], and applied this model to the analysis of previous results obtained from different groups. In those models, the noise source remained in the pump, while signal and idler were coupled to this noise source only through the parametric process. As such, perfectly correlated classical noise between signal, idler, and pump beams would be created, in contradiction to the observed discrepancies [19]. The observation of tripartite entanglement, for instance, would have been easier.

Another model for this excess noise was recently proposed, applied to the study of bipartite entanglement between the fundamental and the second harmonic in parametric conversion [17]. The phase noise introduced is proportional to the field amplitude, and it is presented as Guided Acoustic Wave Brillouin Scattering (GAWBS) [29]. In this case, crystal absorption would be the origin of this extra noise term.

Our model relies on the intrinsic fluctuations of thermal origin in the refractive index, which scatter photons from the carrier into the noise sidebands. Here we extend our model to other experiments. In what follows, based on the cavity parameters (losses, threshold power, free spectral range, detection efficiency) and the measured results of entanglement or squeezing, we calculate the values of the noise coupling terms $\eta_{jk}$. In order to simplify the number of free parameters, we assumed that all the noise coupling scales according to the values independently obtained in our experiment (eqs 21 and 22). Therefore we kept $\eta_{11} = \eta_{22} = \eta_{00}/4$, $\eta_{01} = \eta_{02} = 0.27\eta_{00}$, and $\eta_{12} = 0.16\eta_{00}$, and evaluated the value of $\eta_{00}$ using the model of eq. 14.

Previous measurements of bipartite entanglement in our lab employed another OPO cavity, used in refs. [14, 28]. Noise in these systems could be correctly fitted with $\eta_{00} = 0.72/W$. In the first demonstration of bipartite entanglement in the above-threshold OPO [11], a value $\eta_{00} = 0.64/W$ accounts for the difference of 0.21 units of SQL between the measured quadrature noise and the expected value from the theory without additional noise.

Extending this analysis to other measurements, we carried out the analysis for the frequency-degenerate OPO studied in ref. [10]. In their case, the expected value for the phase quadrature correlation would give a noise level of 47% of the SQL. Instead, they measured excess noise of twice the SQL. A noise coupling $\eta_{00} = 1.16/W$ would explain their results.

Other experiments presented unaccounted sources of excess noise. For instance, in a cascaded $\chi^{(3)}$ process [9], pump squeezing reached only 6% (pump noise equal to 94% of SQL) in the phase quadrature, that could be explained with $\eta_{00} = 0.24/W$. This is a reasonable value, considering here that they used a PPLN (periodically poled lithium niobate) crystal pumped at 1.064 $\mu$m, half the wavelength of the other experiments.

Recently, entanglement between fundamental and second harmonic in process of up or down conversion in a cavity was demonstrated. From the work of ref. [17], us-ing a degenerate OPO/SHG with injection at both wavelengths, we could calculate the relation of the constants $\eta_{kk}$ using the given value for the parameter $\xi_k$, obtaining values of $\eta_{00} = 0.77/W$ and $\eta_{11} = 0.19/W$.

All those experiments, involving diferent crystals, wavelengths and configurations, returned values for the noise coupling of the same magnitude of the results obtained in the present paper. Other experiments with the OPO, such as twin beam generation, involved only amplitude quadratures. Thus, they were free from this excess noise source. The reported values of squeezing matched the expected results once the technical problems were taken in account. The fact that amplitude noise is uncoupled from the fluctuation in the refractive index due to phonons inside the crystal is a good reason why they were observed long before phase quadrature squeezing and entanglement.

On the other hand, the experiment of squeezing in SHG [7] was performed with a scanning cavity that presented squeezing at exact resonance, but very high excess noise for non-zero detuning. In this case, amplitude and phase noise are coupled in the drift matrix, and the excess phase noise owing to the additional phonon noise, expected to be present in that case, was coupled to the amplitude.

For experiments with the OPO below threshold, the pump field acts as a classical field, driving the parametric process, and therefore the noise remains uncoupled. Since these experiments have negligible intracavity power for the generated vacuum modes, the scattering of the central carrier can be neglected. Nevertheless, it may still act as an extra loss source, and may need to be considered when attempting to obtain very high levels of squeezing.

Not all the other experiments considered could be successfully analyzed by our model. Exceptions, resulting in smaller values for the noise coupling $\eta$, were observed. In the article by Peng et al. [12], the difference of 10% of SQL between theory and measurement could be explained by a value of $\eta_{00} = 0.68 \times 10^{-2}/W$, much smaller than the value obtained in the present article. Another article, by Pfister et al. [13], would yield $\eta_{00} = 0.84 \times 10^{-1}/W$ for the difference of $14,2\%$ of the SQL in the phase correlation. In both cases an (almost) open cavity for the pump mode (very low cavity finesse) was employed. This invalidates the assumption of small losses for the pump mode used in the present model.

## VI. CONCLUSION

We analyzed the unexpected phase noise observed in Optical Parametric Oscillators and developed a model that agrees well with the experimental results. The origin of the noise was determined to be the random fluctuations of the refractive index (or the crystal permittivity) induced by thermal phonons within the crystal. This fluctuation scatters light from the mean field into the noise sidebands, acting as an additional noise source in quantum optics measurements.

The noise source was characterized for a cavity with a KTP crystal, and the observed results agree well with the model regarding the noise dependence with power, cavity geometry, wavelength, and temperature dependence. The latter is the strongest evidence of the role played by the phonons. We applied the noise model to the OPO and observed good agreement with the experiment.

Other systems reported in the literature were analyzed, and the model consistently succeded in explaining discrepancies between the theory without extra noise and the observed results. The additional noise value we have inferred for most of the experiments was of the same order of magnitude of the noise we measured in the present paper. This indicates that such an effect may have to be considered in designing other quantum optical systems with nonlinear crystals inside cavities.

Our recent research has been driven by a quest for three-color tripartite pump-signal-idler entanglement in the above-threshold OPO [18, 19]. Previous attempts were frustrated by the presence of this extra noise source of then unknown physical origin. Following the results presented here, by lowering the crystal temperature we finally succeeded in experimentally observing the direct generation of tripartite entanglement [34], producing three entangled fields of different colors. We expect interesting applications in quantum information for this unusual light source. Borrowing a line from Alfred Hitchcock, in *The Trouble With Harry*, we can proudly announce that the trouble with the noise is over.

## VII. ACKNOWLEDGEMENTS

We acknowledge partial support from Conselho Nacional de Desenvolvimento Científico e Tecnológico (CNPq), Coordenação de Aperfeiçoamento de Pessoal de Nível Superior (CAPES), and Deutscher Akademischer Austausch Dienst (DAAD). ML acknowledges support from the Alexander von Humboldt Foundation.

---